\documentclass[
  journal=largetwo,
  manuscript=article-type,
  year=2020,
  volume=37
]{cup-journal}

\usepackage{aas-macros}
\usepackage[utf8]{inputenc}
\usepackage{csquotes}
\usepackage[english]{babel}

\usepackage{amsmath,amssymb}
\usepackage[nopatch]{microtype}
\usepackage{booktabs}
\usepackage{siunitx}
\usepackage{graphicx}
\usepackage{rotating}
\usepackage{multirow}
\usepackage{tablefootnote}
\usepackage{tikz}
\usepackage{upgreek}

\usepackage{biblatex}
\newcommand{\zdm}{{\sc zDM}}
\newcommand{\pccc}{\si{pc\,\centi\metre^{-3}}}

\newcommand{\rubin}{the Rubin Observatory}
\newcommand{\lsst}{LSST}
\newcommand{\single}{single visit}
\newcommand{\singles}{single visits}
\newcommand{\coadds}{10 year co-adds}

\title{Fast Radio Bursts in the Era of the Vera C. Rubin Observatory's Legacy Survey of Space and Time} 

\author{C.~W.~James}
\affiliation{International Centre for Radio Astronomy Research, Curtin University, Bentley, 6102, WA, Australia}
\email[Clancy W.\ James; Kristen Dage]{clancy.james@curtin.edu.au; kristen.dage@curtin.edu.au}

\author{B.~Smith}
\affiliation{International Centre for Radio Astronomy Research, Curtin University, Bentley, 6102, WA, Australia}

\author{K.~Dage}
\affiliation{International Centre for Radio Astronomy Research, Curtin University, Bentley, 6102, WA, Australia}

\author{A.~L.~Chies Santos}
\affiliation{Instituto de F\'isica, Universidade Federal do Rio Grande do Sul (UFRGS), 9500, Porto Alegre, RS, Brazil}

\author{K.~W.~Bannister}
\affiliation{Australia Telescope National Facility, CSIRO Space \& Astronomy, Epping, NSW 1710, Australia}

\author{M.~Caleb}
\affiliation{Sydney Institute for Astronomy, School of Physics, The University of Sydney, Sydney, NSW 2006, Australia}
\alsoaffiliation{ARC Centre of Excellence for Gravitational Wave Discovery (OzGrav), Hawthorn, VIC 3122, Australia}

\author{J.~F.~Crenshaw}
\affiliation{Department of Physics, University of Washington, Seattle, WA 98195, USA}

\author{A.~T.~Deller}
\affiliation{Centre for Astrophysics and Supercomputing, Swinburne University of Technology, Hawthorn, 3122, VIC, Australia}

\author{K.~G.~Lee}
\affiliation{Kavli IPMU (WPI), UTIAS, The University of Tokyo, Kashiwa, Chiba 277-8583, Japan}
\alsoaffiliation{Center for Data-Driven Discovery, Kavli IPMU (WPI), UTIAS, The University of Tokyo, Kashiwa, Chiba 277-8583, Japan}

\author{L.~Marnoch}
\affiliation{School of Mathematical and Physical Sciences, Macquarie University, NSW 2109, Australia}
\alsoaffiliation{Astrophysics and Space Technologies Research Centre, Macquarie University, Sydney, NSW 2109, Australia}
\alsoaffiliation{Australia Telescope National Facility, CSIRO Space \& Astronomy, Epping, NSW 1710, Australia}

\author{K.~M.~Rajwade}
\affiliation{Department of Physics, University of Oxford, Oxford OX1 3RH, UK}

\author{S.~D.~Ryder}
\affiliation{School of Mathematical and Physical Sciences, Macquarie University, NSW 2109, Australia}
\alsoaffiliation{Astrophysics and Space Technologies Research Centre, Macquarie University, Sydney, NSW 2109, Australia}

\author{R.~M.~Shannon}
\affiliation{Centre for Astrophysics and Supercomputing, Swinburne University of Technology, Hawthorn, 3122, VIC, Australia}

\author{B.~Stappers}
\affiliation{Jodrell Bank Centre for Astrophysics, Department of Physics and Astronomy, The University of Manchester, Manchester M13 9PL, UK}

\author{T.~Zhang}
\affiliation{Department of Physics and Astronomy and PITT PACC, University
of Pittsburgh, Pittsburgh, PA 15260, USA}


\begin{document}

\begin{abstract}
Identifying the host galaxies of fast radio bursts (FRBs), and comparing their redshifts and dispersion measures, has unlocked a new probe of the cosmological distribution of ionised gas. However the necessary optical observations to identify FRB hosts, and measure their redshifts, are becoming increasingly onerous as the detection rate of precisely localised FRBs increases. Here we analyse the ability of the Legacy Survey of Space and Time (LSST), being conducted by the Vera C.\ Rubin Observatory, to identify FRB host galaxies, and the utility of LSST photometric redshifts for FRB cosmology.
By combining a model of FRB host galaxy r-band magnitudes, $m_r$, with predictions for the FRB z--DM distribution, we create a method to predict the $m_r(z)$ distribution for the host galaxies of FRBs detected by radio surveys. We then predict these distributions for the coherent modes of the Australian Square Kilometre Array Pathfinder (ASKAP) and MeerKAT. We find that even a \single\ with Rubin will be able to identify 65\% of FRB host galaxies detected by ASKAP's coherent upgrade, `CRACO'; while the final 10~year co-added images will identify 81\% of those from MeerKAT's tied array beams.

We also simulate the impact of using photometric redshifts for a simplified analysis to determine $H_0$, finding that estimated photo-z errors result in a decreased precision of only 7\% on $H_0$ for ASKAP's CRACO system. The impact of missing faint FRB hosts, which are likely at higher redshifts, is more significant, and might degrade sensitivity to $H_0$ by 47\%, or 62\% when combined with photo-z errors.
All told, Rubin's LSST will be an incredibly powerful survey for facilitating FRB cosmology, although supplemental observations may be useful for particularly faint and distant host galaxies.
\end{abstract}

\section{Introduction}
\label{sec:intro}

Fast Radio Bursts (FRBs), mysterious high energy radio transients that last for less than a few milliseconds, have been an ongoing enigma since their discovery in 2007 \citep{lorimer_bright_2007}. FRBs trace the cosmological distribution of ionised gas through their dispersion measure \citep[DM;][]{Macquart2020}. This has enabled FRBs to be used to constrain (for instance) the Hubble constant \citep{WuHubble22,HagstotzHubble22,2022MNRAS.516.4862J}, fluctuations in the cosmological matter density \citep{2024ApJ...965...57B}, the fraction of ionised gas residing in the circumgalactic and intergalactic media  \citep{FLIMFLAMdr1,2025NatAs...9.1226C}, and the gas in the Milky Way's halo \citep{2023ApJ...946...58C,2025AJ....169..330R,HoffmannHalo26}.

Studies of FRB host galaxies can also be used to shed light on their progenitors. The hosts of apparently non-repeating FRBs tend to follow the star-forming main sequence \citep{gordon_demographics_2023,2025ApJ...991...85L}, while the small fraction of repeating FRBs \citep[$\lesssim3$\%][]{2026ApJS..283...34C} prefer star-forming environments such as dwarf galaxies and outer spiral arms \citep[e.g.,][]{2017Natur.541...58C,marcote_repeating_2020,Niu2022,2026ApJ...996L..16M}. The role of metallicity in determining FRB host galaxies is currently debated \citep{2024Natur.635...61S,2026ApJ...997L...6Y}, although typically low-metallicity environments, such as the hosts of superluminous supernova (SLSNe) and long-duration gamma-ray bursts (LGRBs), seem to be disfavoured \citep{Bhandari+22}. It has also been argued that the main inconsistency with a star-forming origin is an excess of low-magnitude hosts with little star-formation \citep{2020ApJ...905L..30S}. Spurred by the localisation of FRB\,20200120E to a globular cluster (GC) in M81 \citep{2022Natur.602..585K}, GCs have also been suggested as a tracer of FRB hosts \citep{2026ApJ...996...78H}.

Key to these discoveries is the radio localisation of FRBs to (sub-)arcsecond precision, the subsequent identification of their host galaxies by optical observations, and measurement of their redshifts \citep{2017Natur.541...58C,121102Host,bannister_single_2019}. While a large number of unlocalised FRBs absolutely can be used for cosmological inference \citep[e.g., ][]{2025arXiv250608932W}, the utility of a FRB with known redshift is far greater than that of one with DM information only, due to intrinsic fluctuations in the DM budget \citep{2022MNRAS.516.4862J}. This has resulted in a huge global effort to increase the number of FRBs with arcsecond-scale radio localisations, in particular the more TRANsients and Pulsars (MeerTRAP) survey on MeerKAT \citep{2025MNRAS.tmp.2025P}; the Commensal Real-time ASKAP Fast Transients (CRAFT) survey on the Australian Square Kilometre Array Pathfinder \citep{Shannon_ICS}; the Deep Synoptic Array \citep[DSA-110;][]{2024Natur.635...61S}, and the Canadian Hydrogen Intensity Mapping Experiment's Fast Radio Burst group and associated outriggers \citep[CHIME/FRB; ][]{2025ApJS..280....6C}.

Most FRB observations are either commensal, or performed by transit telescopes, meaning that FRB discoveries are spread over a large area of the sky. Identification of FRB host galaxies therefore require wide-field optical surveys, such as SDSS \citep{2022ApJS..259...35A}, PanSTARRS \citep{2010SPIE.7733E..0EK,2016arXiv161205560C}, and DECam \citep{2015AJ....150..150F}. However, the relatively shallow depths of these surveys (magnitude limits of approximately $m_r=22.7$, $m_r=23.3$, and $m_r=23.27$ respectively)
leads to incompleteness in the resulting FRB host catalogues, affecting their use for cosmological parameter estimation \citep{2025PASA...42...17H}. This incompleteness has been partially addressed with dedicated 8-m class observations, e.g., the sample presented by \citet{Shannon_ICS}. However,
the increasing number of precisely localised FRBs has put severe constraints on the time required to identify FRB hosts that are too faint to be found in existing catalogues, and perform the spectroscopy to identify their redshifts. Furthermore, as FRB-hunting radio telescopes become more sensitive, they are finding ever-more-distant FRBs, often requiring space-based observations to find a host \citep{Marnoch2023,2025arXiv250801648C}. The next generation of telescopes, such as CHORD \citep{CHORD} and DSA-2000 \citep{DSA2000}, are expected to find $\mathcal{O}\sim 10,000$ FRBs/year, making dedicated spectroscopic follow-up of their hosts completely infeasible.

This situation is expected to change with the advent of the Vera C. Rubin Observatory's Legacy Survey of Space and Time (LSST; \citealt{2022ApJS..258....1B}). Its 3 billion-pixel digital camera will survey the entire Southern Hemisphere ---  the same sky scanned in radio by leading FRB detection telescopes MeerKAT, ASKAP, and, eventually, SKA-Mid and SKA-Low. In this paper, we first discuss the prospects for identifying the host galaxies of FRBs detected by ASKAP and MeerTRAP with the Rubin Observatory, and then explore the impact that uncertainties in current photometric redshift techniques may have on FRB cosmological parameter estimation.

\section{How many host galaxies?}
\label{sec:frbs}

In this Section, we determine the fraction of FRB host galaxies that \rubin\ is expected to be able to observe. We first determine appropriate optical magnitude limits for \rubin, and then build a simple model of the intrinsic FRB host galaxy magnitude distribution. Finally, we use a model of FRB observations with ASKAP and MeerTRAP to determine redshift, and, hence, magnitude distributions for FRB hosts from these surveys, and determine the fraction visible to \rubin. Throughout, we characterise the detectability of FRB hosts via their r-band magnitudes $m_r$, which have commonly been used for host identification in follow-up observations \citep[e.g.][]{Shannon_ICS}, and to assess the statistical probability of host association \citep{PATH}.

\subsection{What's visible to Rubin?}
\label{sec:fractions}

The Rubin Observatory achieved First Look in mid-2025, and the LSST has now begun in 2026. Data Preview~1 (DP1), based on commissioning camera data, has already been released, with Data Release 1 planned for 2028.\footnote{See \url{https://rubinobservatory.org/for-scientists/resources/early-science}.} In addition to the full survey of the Southern Sky, as well as the North Ecliptic Spur and part of Virgo, Rubin will also target specific `deep-drilling' fields which will be observed at a higher cadence\footnote{\url{https://survey-strategy.lsst.io/baseline/}}. 

Rubin observes with six filters, $ugrizy$, which will be individually co-added over the full 10 year survey. The 5 $\sigma$ limiting magnitudes for both \singles\ and \coadds\ can be found in \citet{2022ApJS..258....1B}, but we specifically highlight that the r-band single filter limiting magnitude (5$\sigma$) for a single visit is $m_r^{\rm lim}=24.7$, and $m_r^{\rm lim}=27.5$ in the \coadds.  We therefore use these two magnitude limits going forward as characteristic minimal and maximal sensitivities of \rubin.

\subsection{Host galaxy magnitudes}
\label{sec:hosts}

Numerous studies have analysed FRB host galaxies, focusing on star-formation rate, total stellar mass, and various other properties \citep[e.g.][]{2020ApJ...903..152H,Bhandari+22,gordon_demographics_2023,2024Natur.635...61S}. However, the only study we are aware of that models an intrinsic distribution of FRB host galaxy magnitudes, and their redshift dependence, is \citet{Marnoch2023}. In this work, the authors use 23 FRB host galaxies with optical spectra, and model the expected r-band and $K_s$-band magnitudes were those galaxies to have been located at redshifts in the range 0--2. This study however does not account for galaxy evolution (the sample is based on hosts detected with $z \lesssim 0.5$), nor the bias due to FRBs originating in hosts too faint to be detected, and thus the results we derive here will also be subject to these deficiencies. However, it is a plausible first approximation given the level of understanding of the FRB host population in the current literature. It bears noting that the combination of \lsst\ data with numerous low-redshift FRBs from (in particular) ASKAP will help address the latter of those two limitations, as discussed in Section~\ref{sec:lsstdetectionfraction}. Extending the work of \citet{Marnoch2023} with the increasingly large sample of FRB host galaxies would also be of great benefit --- however, spectroscopic data for the vast majority of surveys is either unavailable, or complete to significantly brighter magnitudes and/or lower redshifts \citep{2024Natur.635...61S,2025MNRAS.tmp.2025P}. We hope to include the recent results of \citet{2026ApJ..1001..118M} for 12 new FRB hosts in the future.

\begin{figure}
    \centering
    \includegraphics[width=\columnwidth]{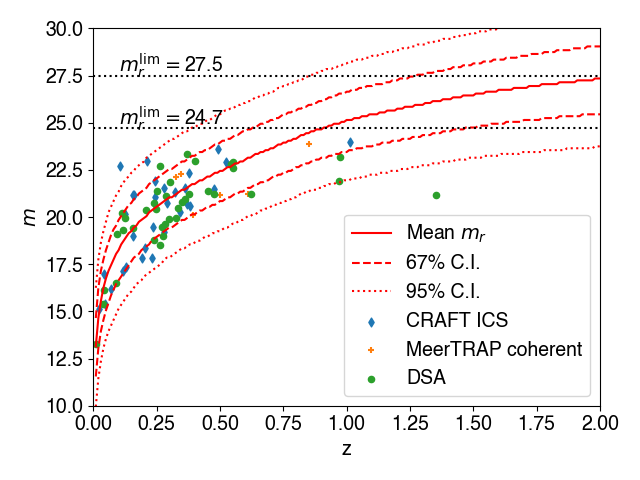}
    \caption{Mean, and 67\% ($\pm 1 \sigma$) and 95\% confidence ($\pm 2 \sigma$) intervals, of the FRB r-band host galaxy magnitude distribution, $m_r$, as a function of redshift, using the FRB host galaxies analysed by \citet{Marnoch2023}. Also shown for comparison are the $m_r$ magnitude limits from \lsst\ \single\ and \coadds, and FRB host galaxies (measured using a variety of filters) detected in CRAFT ICS \citep{Shannon_ICS}, MeerTRAP coherent \citep{2025MNRAS.tmp.2025P}, and DSA \citep{2024Natur.635...61S,2025NatAs...9.1226C} observations.}
    \label{fig:mean_rms}
\end{figure}

\begin{figure}
    \centering
    \includegraphics[width=\linewidth]{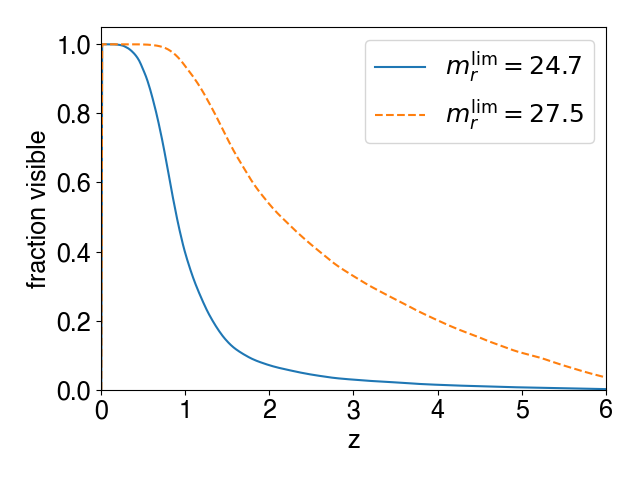}
    \caption{Fraction of FRB hosts visible to \rubin\ for optical r-band limits corresponding to \single\ and \coadds\ observations.}
    \label{fig:fractions}
\end{figure}

Both \citet{Marnoch2023} and \citet{2025arXiv250801648C} have used these data by effectively treating the $m(z)$ distribution as a series of $\delta$-functions at the calculated values of the 23 hosts, so that the probability $p(m_r < m_r^{\rm lim})$ can be obtained by counting the fraction $N(z)/N_{\rm tot}$ of hosts with estimated magnitudes less than $m_r^{\rm lim}$, i.e.\ it becomes a step-function in $m_r$. To avoid this step-like behaviour, we evaluate the mean $\mu_r(z)$ and standard deviation $\sigma_r(z)$ of the FRB host galaxies studied by \citet{Marnoch2023}, and assume an underlying normal distribution, i.e.\ at any $z$, $m_r \sim N(\mu_r,\sigma_r)$. These values are shown in Figure~\ref{fig:mean_rms}. The limit from a \single\ (\coadds) corresponds approximately to the mean FRB host magnitude at $z \sim 1$ ($z \sim 2$). Alternatively, we can also calculate the fraction of FRB hosts visible as a function of redshift. This is given in Figure~\ref{fig:fractions}. Note that we will reach $\sim$50\% completeness at cosmic noon for the 10\,yr LSST co-addition.

Since they can be targeted with follow-up radio observations, a large fraction of repeating FRBs have been localised to their hosts \citep[e.g.][]{2017Natur.541...58C}. However, these may come from a different host population \citep{Bhandari+22,2022Natur.602..585K}, and likely are only observed to repeat because they are nearby \citep{Gardinier2021_frbpoppy_repeaters,2023PASA...40...57J}. Furthermore, publication bias --- whereby unique or unexpected results are preferentially published ahead of standard results --- can lead to particularly high-redshift FRBs being published ahead of other bursts detected in the same period by the same instrument, with examples being \citet{2023Sci...382..294R} and \citet{2025arXiv250801648C}. The most statistically complete catalogues of FRB host galaxies come from interferometers, which are able to routinely localise once-off FRBs to arcsecond precision, enabling a robust host identification \citep{2017ApJ...849..162E}.

The largest such catalogues are currently those from ASKAP \citep{Shannon_ICS}, MeerTRAP \citep{2025MNRAS.tmp.2025P}, and DSA \citep{2024Natur.635...61S,2025NatAs...9.1226C}. While 17 of the 23 FRBs used by \citet{Marnoch2023} came from the ASKAP catalogue, the DSA and MeerTRAP catalogues are independent, allowing these to be used to test the reliability of our model. Comparing these host galaxies to our model of $m_r(z)$ in Figure~\ref{fig:mean_rms}, it is clear that these predictions have held for all FRB host galaxies, with the exception of FRB\,20230521B at $z\sim1.3$, the redshift of which was identified only via a single emission line \citep{2025NatAs...9.1226C}. There also appears to be a magnitude limit of $m_r \sim 24$, above which no FRB hosts are confidently identified by the typically ground-based, 8\,m class optical telescope observations used to identify them. We partially attribute this to the inability of optical observations to observe the host, and primarily to the difficulty in uniquely identifying it amongst field galaxies. Indeed, B.~Anderson et al.\ (in prep, 2026) has shown that, with standard statistical methods \citep{PATH}, requiring a posterior likelihood of at least 90\% results in significant bias against hosts with $m_r>22$. This similarly suggests that many FRB host catalogues are incomplete above a redshift of $\sim 0.2$.



\subsection{Simulating FRB Observations}
\label{sec:simulations}

To estimate the redshift (and hence, magnitude) distribution of FRB host galaxies, we consider FRB observations by ASKAP, MeerKAT, and SKA-Mid. While \lsst\ will have some overlap with the sky viewed by CHIME, and in the future, DSA-2000 \citep{DSA2000} and CHORD \citep{CHORD}, the majority of FRBs discovered by these instruments will not lie within the \lsst\ sky coverage.

For ASKAP, we simulate observations with the CRAFT Coherent Upgrade \citep[CRACO; ][]{CRACO}, using a time resolution of 13.8\,ms as per recent observations (Yuanming Wang et al., in prep., 2026). Rather than iterate over all ASKAP observing configurations, we choose a characteristic central frequency of 1300\,MHz, 280\,MHz real-time bandwidth, and use the {\sc square\_6x6} beam pattern with a beam separation of 0.9\,deg. Assuming 25 ASKAP antennas are used for real-time imaging, we estimate a fluence threshold of 3.7\,Jy\,ms to a burst of duration less than 13.8\,ms detected at boresight. We simulate MeerKAT as per \citet{2025arXiv250801648C}, using the coherent (tied array beam) mode, with a mean frequency of 1284\,MHz, yielding a threshold of 0.069\,Jy\,ms to a 1\,ms FRB \citep{2023MNRAS.524.4275J}. We note that the estimated FRB detection threshold of the AA4 configuration of SKA-Mid in band 2 (L-band) is 0.059\,Jy\,ms, similar to that of MeerKAT (FRB cosmology --- SKA science chapter; L.~Spitler, E.~Keane et al., eds., in prep., 2026). Therefore, we expect that all the results we present here for MeerKAT will likely also apply to that instrument.

\begin{figure}
    \centering
    \includegraphics[width=\linewidth]{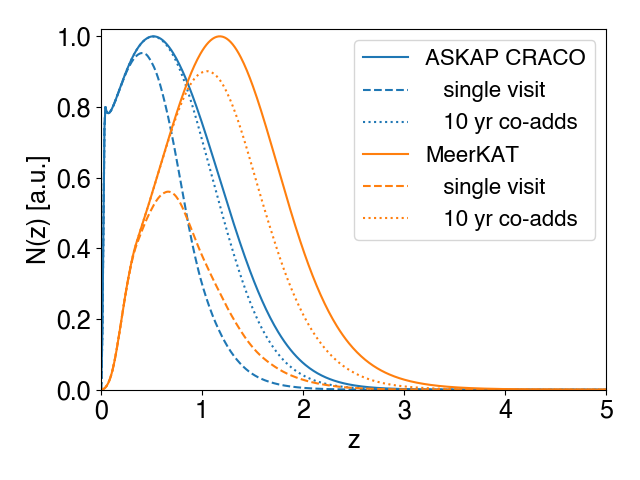}
    \caption{Redshift distribution of FRBs, normalised to a peak of unity, for FRBs detected by ASKAP's CRACO system, and MeerKAT's coherent MeerTRAP mode. Also shown are the redshift distributions of FRB host galaxies accessible with a single LSST visit (dashed) and 10yr co-adds  (dotted).}
    \label{fig:zdists}
\end{figure}

We use standard Planck cosmology \citep{planck_collaboration_planck_2016}, and take FRB population parameters from \citet{2025PASA...42...17H}. The resulting FRB redshift distributions are shown as the solid curves in Figure~\ref{fig:zdists}, normalised to a peak of unity.

\subsection{Fraction of FRB hosts detectable by \lsst}
\label{sec:lsstdetectionfraction}

\begin{table}[]
    \centering
    \begin{tabular}{c|cc}
    Telescope & $f_{24.7}$ & $f_{27.5}$ \\
    \hline
     ASKAP/CRACO    & 0.65 & 0.93 \\
     MeerTRAP coherent & 0.37 & 0.81
    \end{tabular}
    \caption{Fractions of FRB hosts detected in two FRB surveys estimated to be visible at \lsst\ \single\ and \coadds\ r-band magnitude limits of $m_r=24.7$ and $m_r=27.5$, respectively.}
    \label{tab:fractions}
\end{table}

\begin{figure}
    \centering
    \includegraphics[width=\linewidth]{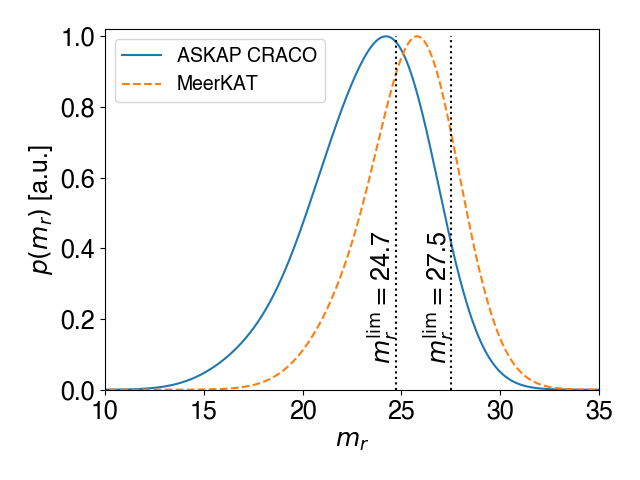}
    \caption{$r$-band magnitude distribution of FRB host galaxies expected to be detected by ASKAP's CRACO system, and MeerKAT. Also shown (as vertical dotted lines) for comparison are the magnitude limits from \lsst\ \singles\ and \coadds.}
    \label{fig:mrdists}
\end{figure}

We can now evaluate, for any given survey, the total magnitude dependence of the FRB host galaxy distribution, by convolving the redshift distributions of Figure~\ref{fig:zdists} with the detectable host fractions of Figure~\ref{fig:fractions}. The resulting distributions of $m_r$ are given in Figure~\ref{fig:mrdists}. Coincidentally, the predicted peak in the magnitude distribution of ASKAP CRACO FRB hosts almost coincides with the \lsst\ \single\ magnitude limit, resulting in 65\% of hosts being visible, while the $m_r=27.5$ limit for \coadds\ is above the predicted peak for MeerTRAP FRB hosts, with 81\% of hosts predicted to be detectable. These fractions are summarised in Table~\ref{tab:fractions}.

We also evaluate the redshift-dependent fraction of FRBs visible to each survey given the magnitude limits of \lsst\ \singles\ and \coadds\ --- these modified redshift distributions are also shown in Figure~\ref{fig:zdists}.

Comparing Figure~\ref{fig:zdists} to Figure~\ref{fig:fractions} also makes it clear that the majority of ASKAP CRACO FRBs are predicted to be detected from a redshift range in which every FRB host galaxy should be visible in the \lsst\ \coadds, under our current understanding of the FRB host galaxy population. This implies that any limitations in the current models of the intrinsic magnitude distribution of FRB host galaxies (at least at low redshift) would be highlighted by the combination of \lsst\ and CRACO, enabling shortcomings to be quantified and addressed.

\subsection{Follow-up strategies}
\label{sec:followup}

\begin{figure}
    \centering
    \includegraphics[width=\linewidth]{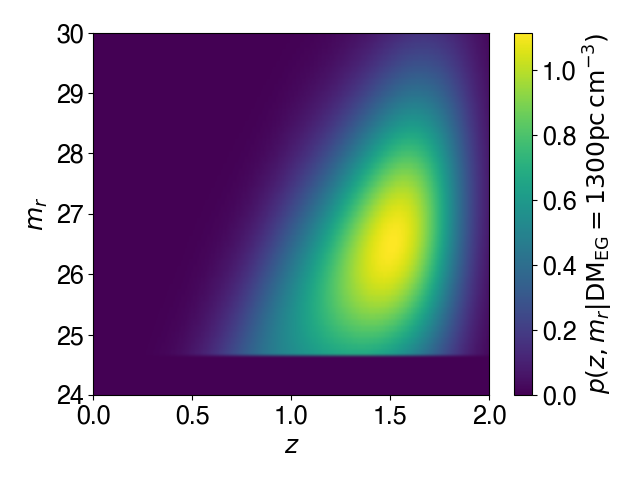}
    \caption{Probability distribution $P(z,m_r|{\rm DM}_{\rm EG} = 1300$\,\pccc$)$ for an FRB detected by CRACO, assuming the true host is fainter than the Rubin single visit magnitude limit of $24.7$.}
    \label{fig:pmzgu}
\end{figure}

Our predictions for the redshift--magnitude distribution of FRB host galaxies also allow us to make predictions for this distribution in the event that the host galaxy of a particular FRB is undetected. This can then be used to guide follow-up observations. Combining this distribution with the detection sensitivity and available observation time of the follow-up instrument would determine the probability of the host follow-up observation being successful, while the specific science case being analysed would determine the relative importance of finding a particular host galaxy of a given magnitude and redhsift. For instance, FRB host galaxy studies may give equal weighting to each host galaxy, while studies aimed at determining $H_0$ (a toy example of which is presented in \S\,\ref{sec:parameters}) may value high-redshift bursts more than low-redshift bursts. Rather than attempting to optimise for a particular science case, we illustrate the joint probability $P(z,m_r)$ in the example case of an FRB with ${\rm DM}_{\rm EG} = 1300$\,\pccc\ detected with ASKAP/CRACO in Figure~\ref{fig:pmzgu}. Such distributions are specific for each FRB DM, and each follow-up instrument, but can be readily calculated using the \zdm\ code. They also have a similar general shape to that given in Figure~\ref{fig:pmzgu}. Projections onto the $m_r$ or $z$ axes are trivial to calculate, and can be used to estimate the likely redshift range of FRBs with unseen hosts, as performed by \citet{Marnoch2023}.

\subsection{Complementarity of other surveys}

Several other wide-field surveys are being planned or conducted on comparable timescales to LSST. The Euclid Wide Survey will cover $\sim 14,500$\,deg$^2$ --- most of the sky, except low Galactic and ecliptic latitudes, spread approximately evenly between the Northern and Southern Hemispheres. It will have an AB-limiting magnitude of $m_{\rm AB} = 26.2$, although survey images of a characteristic extended source, with $10\,\sigma$ detection, require $m_{\rm AB} \le 24.5$ \citep{2022A&A...662A.112E}. Euclid photometric redshifts are expected to be more accurate than that of LSST, producing $\sigma_z \le 0.002 (1+z)$ \citep{2025A&A...702A.155D}. This will be the most relevant survey for identifying FRB host galaxies detected by Northern-hemisphere instruments such as CHORD and DSA.

Several spectroscopic surveys with relatively wide fields are also being planned. The Nancy Grace Roman Space Telescope's 2,400\,deg$^2$ High-Latitude Wide-Area Spectroscopic Survey, with additional $H$-band data over $2,700$\,deg$^2$, will provide redshifts with an accuracy of $\sigma_z < 0.001 (1+z)$ for galaxies with magnitudes $m_{\rm AB} \lesssim 24.0$ (depending on the filter), almost exclusively in the Southern Hemisphere \citep{2022ApJ...928....1W}.\footnote{See also \url{https://science.nasa.gov/mission/roman-space-telescope/}} WAVES-Wide will cover 1,150\,deg$^2$, split between one equatorial and one Southern field, for galaxies with $m_Z \le 21.2$, and a likely redshift of $z < 0.2$ \citep{2019Msngr.175...46D}. Both will prove crucial for establishing a set of FRBs with high-fidelity redshift measurements against which Rubin data can be calibrated --- and providing additional information, such as star-formation histories, which may be important for understanding FRB origins \citep{gordon_demographics_2023}.

The degree to which these latter surveys will overlap with detected FRBs depends significantly on the observation strategies of FRB-hunting instruments. Deep radio surveys naturally target regions with correspondingly deep optical catalogues, producing an excess of FRBs with high-quality optical data. Current examples include the Deep Investigation of Neutral Gas Origins \citep[DINGO; ][]{DINGO} survey with ASKAP, and the Looking At the Distant Universe with the
MeerKAT Array \citep[LADUMA; ][]{LADUMA} survey --- both deep HI surveys that target the GAMA-23 field for 3,200\,hr, and the Chandra Deep Field South for 5,000\,hr, respectively.

In this work, we do not explore the full parameter space of different FRB-hunting instruments, and the time at which each spends observing (and hence, discovering FRBs) in the various fields of existing and planned optical surveys. Neither have we considered Galactic extinction, which not only degrades optical images, but is correlated with Galactic scattering, which reduces the FRB rate at low latitudes \citep{2026ApJS..283...34C}, and uncertainties in the Galactic DM budget \citep[see e.g.,][]{2025AJ....169..330R}, which reduces the utility of low-latitude FRBs for analysing the cosmic distribution of ionised gas \citep{HoffmannHalo26}. We suggest rather that analyses such as this one be performed when considering specific future proposals.

\section{Parameter estimation with photo-z}
\label{sec:parameters}

The majority of cosmological studies with FRB host galaxies utilise spectroscopic redshifts. However, Rubin will provide only photometric redshifts, the estimation of which is currently in development \citep[e.g.][]{2026OJAp....958200R} --- 
however, initial studies with the DP1 data are very promising,\footnote{\url{https://sitcomtn-154.lsst.io/}\label{footnote}} although due to a limited training sample, the photometric redshift estimates are not robust past z$\sim$1.5, and we further caution that the magnitude distribution of the DP1 photo-z test is not representative of the full LSST magnitude-limited sample, and due to the limited data volume, the test set was not reweighted to match the test set to the LSST distribution \citep{2025arXiv251007370Z}. In this DP1 study, six different algorithms for determining photo-zs from Rubin DP1 data, covering the Extended Chandra Deep Field South, are tested against spec-zs. All algorithms tested had broadly similar performance, generating approximately 10–20\% outliers with large ($\delta$z / (1+z) > 0.15) redshift errors \citep{2025arXiv251007370Z}. Here, we arbitrarily use a k nearest neighbour (kNN) algorithm \citep{2026OJAp....958200R}, which showed a characteristic rms error in $z_{\rm photo}-z_{\rm spec}$ of $\sigma_z=0.035$ when excluding outliers, as a typical value. This error is consistent with other studies, e.g., \cite{2018PASJ...70S...9T}. However, we note that DP1 has particularly shallow $u$ and $y$ bands, and the increased depth of these bands in full survey should reduce the number of outliers.

To determine the effect of such errors on cosmological parameter estimation, we first generate a mock FRB sample, and then use this to estimate Hubble's Constant, $H_0$. While this is an overly simplified analysis, we aim only to get a characteristic order-of-magnitude estimate in the uncertainty due to the use of photo-zs rather than spec-zs. In particular, we assume that the magnitude threshold required to estimate a photometric redshift is identical to that required to identify a host galaxy. In reality, it is likely that faint galaxies may not be detected in Rubin's $u$-band, increasing $\sigma_z$ for that subset --- thus, our analysis should be updated to reflect this once photometric redshift determination with LSST is more mature.

\subsection{Monte Carlo simulation of photometric redshifts}
\label{sec:MC}

\begin{figure}
    \centering
    \includegraphics[width=\linewidth]{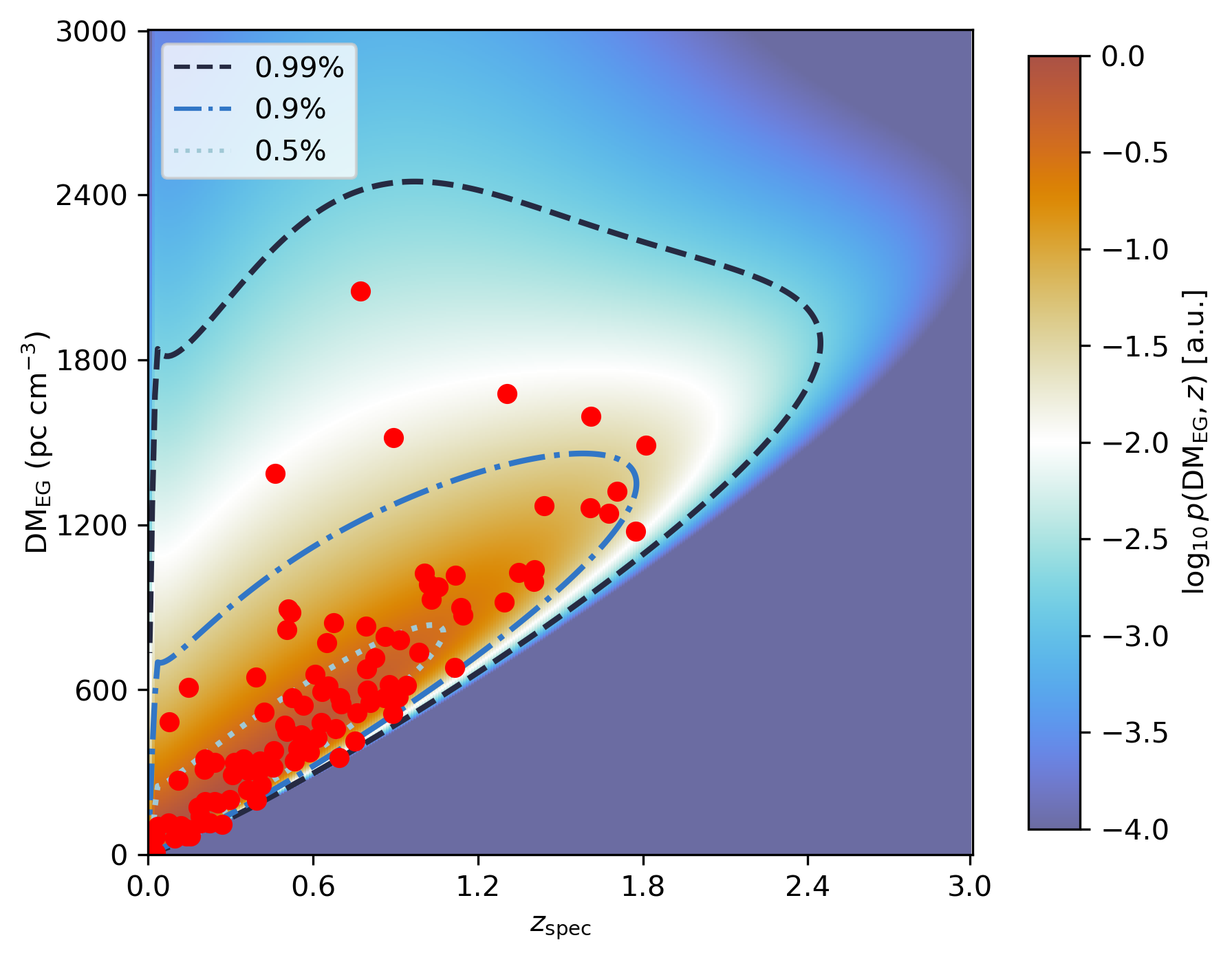}
    \includegraphics[width=\linewidth]{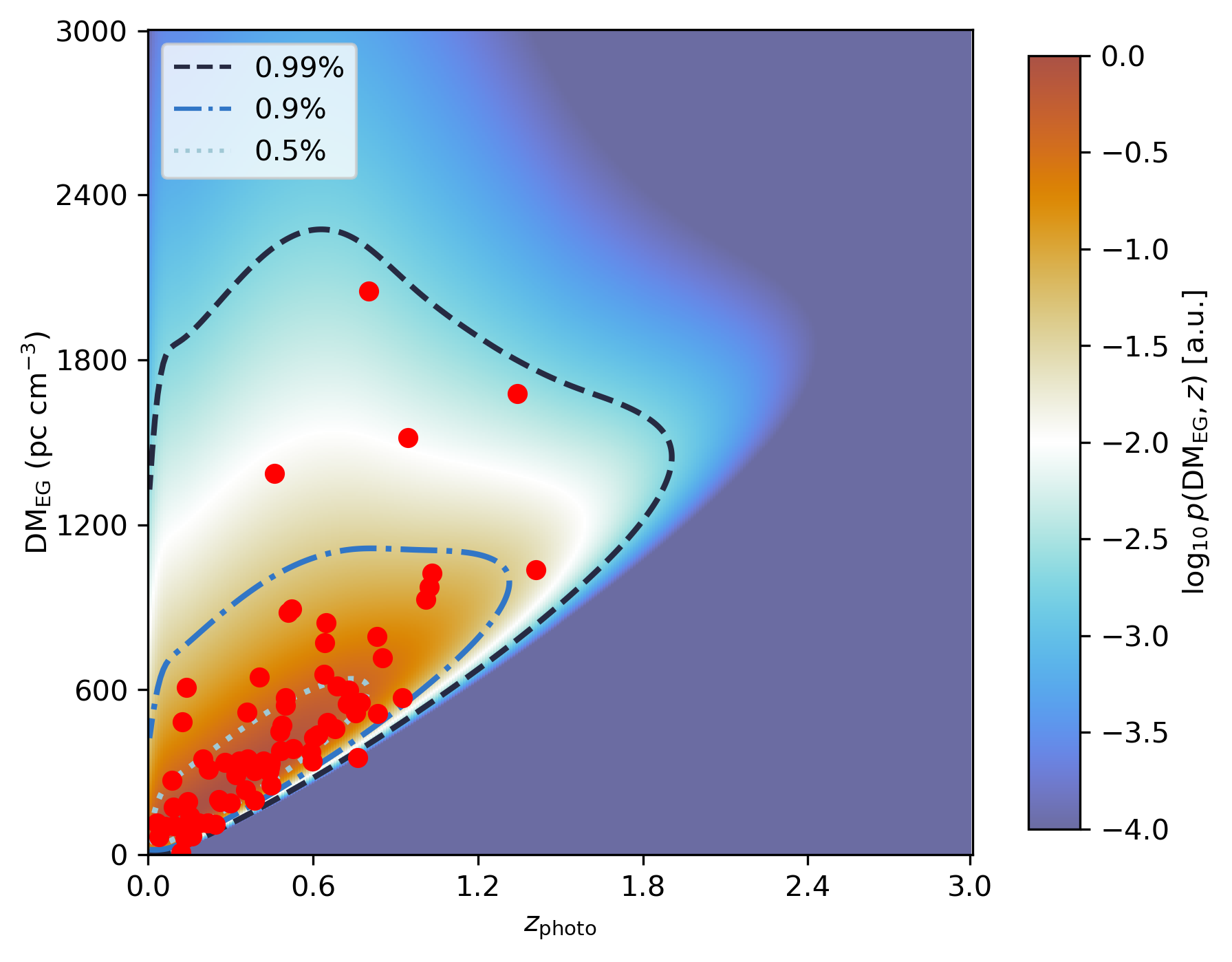}
    \caption{\label{fig:MC} zDM grids and their Monte Carlo FRBs for different LSST parameters: all CRACO FRB host galaxies (top), and including both photo-z errors $\sigma_z$ and host galaxy magnitude limits $m_r^{\rm lim}$ (bottom). The shading indicates the probability density of the simulated truth distribution from which FRBs are sampled, modified in the lower plot by LSST observational effects, while the points indicate the actual sampled FRBs used in this study.}
\end{figure}

\begin{figure}
    \centering
    \includegraphics[width=\linewidth]{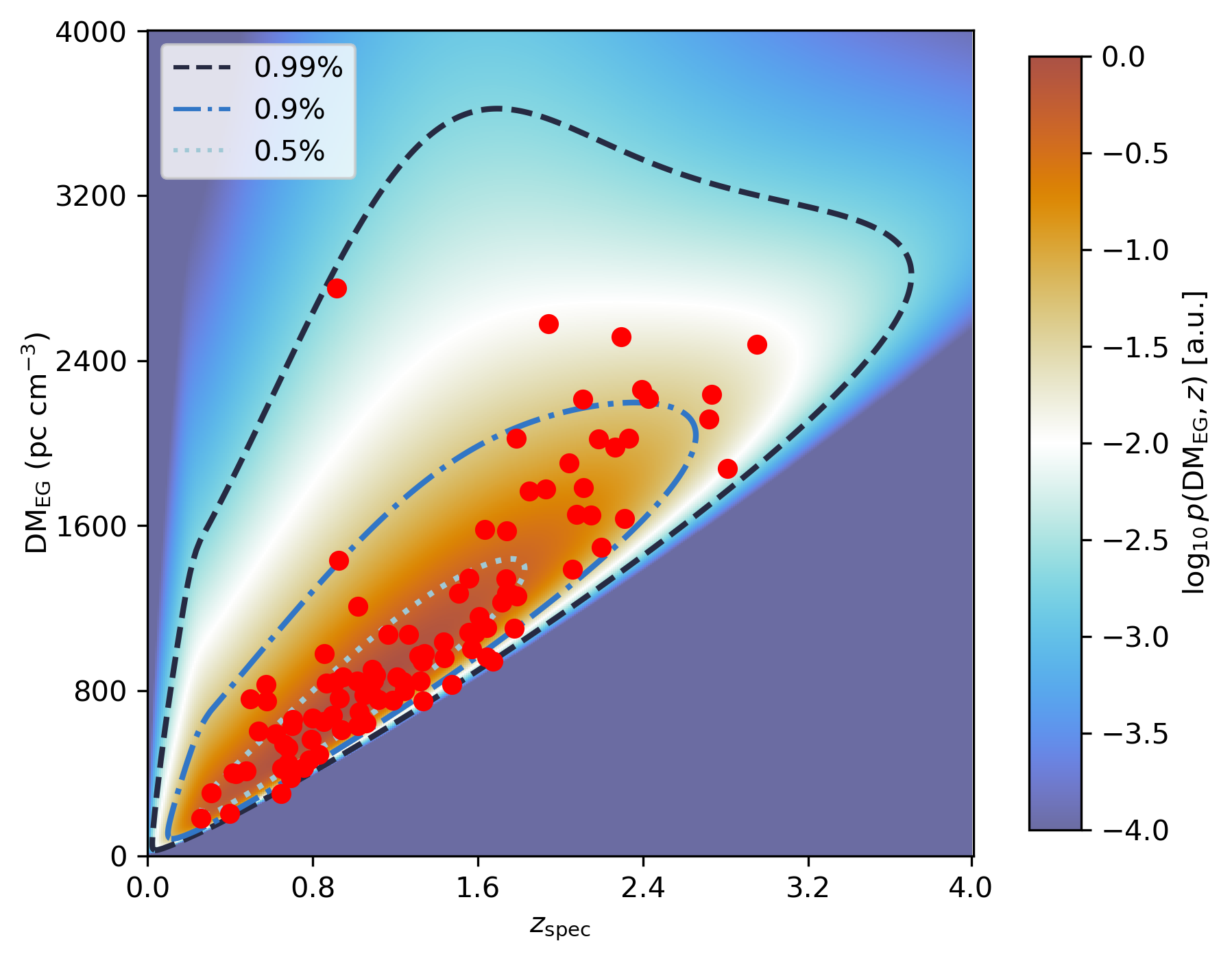}
    \includegraphics[width=\linewidth]{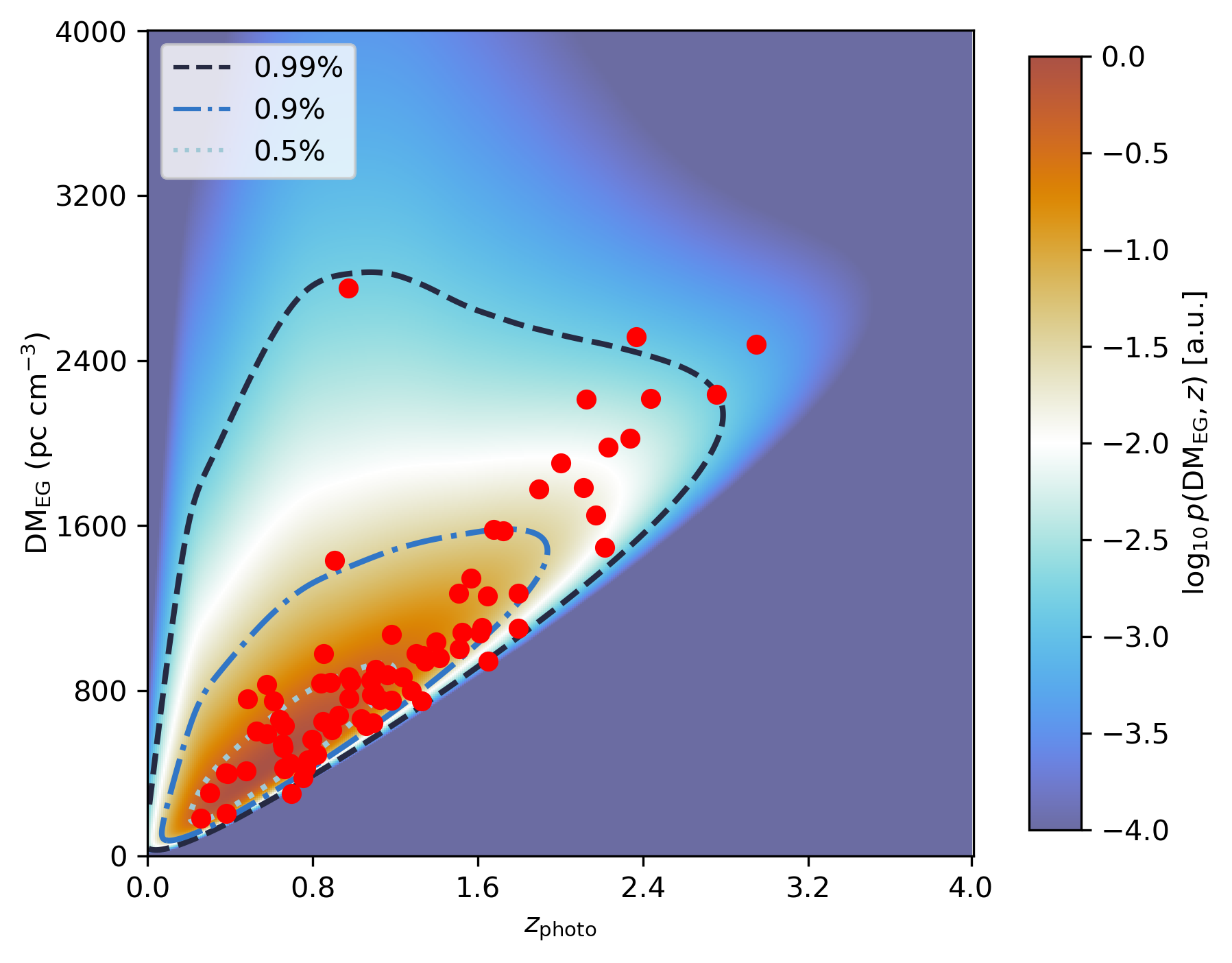}
    \caption{\label{fig:MC2} As per Figure~\ref{fig:MC}, but for MeerKAT coherent, and $m_r^{\rm lim}=27.5$.}
\end{figure}

To model errors in photometric redshift, we assume Gaussian errors with a standard deviation, $\sigma_z$, of $0.035$, as per the kNN algorithm\textsuperscript{\ref{footnote}}.  We also consider the fraction of visible hosts using $m_r^{\rm lim}= 24.7$ from Figure~\ref{fig:fractions}. These effects were implemented into the simulation of ASKAP's coherent (`CRACO') observation mode from \citet{CRACO} for 900\,MHz observations described in \S\,\ref{sec:simulations} to produce modified joint detection probabilities of redshift and dispersion measure, $p(z,{\rm DM})$ (the redshift distributions shown in Figure~\ref{fig:zdists} are simply these joint $p(z,{\rm DM})$ distributions, summed over the DM axis). A FRB survey of 100 FRBs was generated through Monte Carlo sampling of $p(z,{\rm DM})$, both with and without errors from photo-z estimation, and with and without limiting to the observable $m_r^{\rm lim}$. This Monte Carlo procedure was first used to estimate the sensitivity of the derived value of $H_0$ to CRACO by \citet{2022MNRAS.516.4862J}. Examples of these grids for the unmodified case, and including both $\sigma_z$ and $m_r^{\rm lim}=24.7$, are given in Figure~\ref{fig:MC}. We have also repeated the exercise for MeerTRAP coherent mode, again using the parameters described in \S\,\ref{sec:simulations}, using $m_r^{\rm lim} = 27.5$ --- these examples are shown in Figure~\ref{fig:MC2}.

\begin{figure}
\centering
\includegraphics[width=\linewidth]{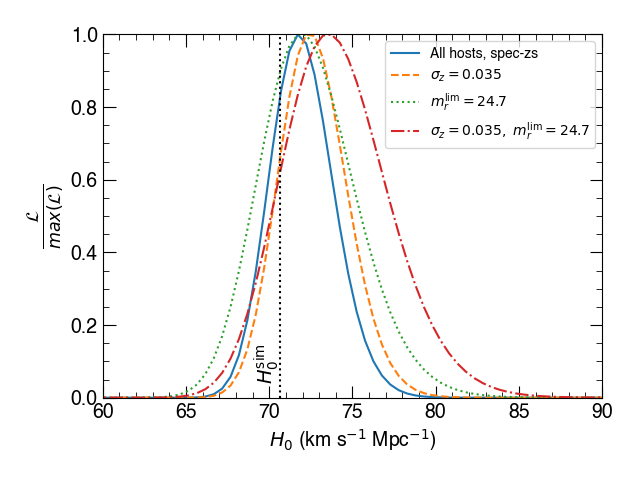}
\caption{Normalised scan of likelihoods of $H_{0}$ for 100 simulated FRB detections by ASKAP CRACO, including limitations of LSST photometric $z$ estimates ($\sigma_z = 0.035$); magnitude limit of $m_r^{\rm lim}=24.7$ on host observations; and both together.}
\label{fig:H0}
\end{figure}

We test the sensitivity of $H_0$ estimation to these photometric adjustments by generating zDM grids for $H_{0}$ in 91 uniformly spaced increments between 60 and 90\,km\,s$^{-1}$\,Mpc$^{-1}$, and evaluating the likelihood of each fake survey to the resulting predicted z--DM distribution. The likelihood contains terms giving the probability for the total number of FRBs observed compared to expectations, their positions in z--DM space, and the ratio of signal-to-noise to threshold signal-to-noise given that position in the z--DM grid --- they are calculated according to \citet{james_zdm_2022}. Exactly the same Monte Carlo sample was used for all likelihood scans, except for the FRBs removed due to the simulated magnitude limit, so that Monte Carlo variation did not impact the relative precision of $H_0$ estimation.
The resulting likelihoods for all combinations of including/excluding $\sigma_z$ and $m_r^{\rm lim}$ effects are given in Figure~\ref{fig:H0}, in the case of CRACO. We characterise the sensitivity to $H_{0}$ via the full-width half-max of the likelihood function, which is given in Table~\ref{tab:H0}. We note that the accuracy of the measurements are determined by the random sampling of FRBs, and are correlated between all four simulations for the one instrument (CRACO or MeerTRAP), since exactly the same FRB sample was used.

\begin{table}[]
    \centering
    \begin{tabular}{r|cc|cc}
                    & \multicolumn{2}{c|}{ASKAP/CRACO} & \multicolumn{2}{c|}{MeerTRAP coherent} \\ 
         & FWHM  & Increase & FWHM  & Increase \\
        Parameters & km\,s$^{-1}$\,Mpc$^{-1}$ & \% & km\,s$^{-1}$\,Mpc$^{-1}$ & \% \\
        \hline
        All, spec-zs &  4.44 & - & 3.14 & -\\
        $\sigma_z=0.035$ & 4.74 & 6.8 & 3.25 & 3.3\\
        $m_r^{\rm lim}$ & 6.51 & 46.8 & 3.76 & 19.6 \\
        $\sigma_z$ \& $m_r^{\rm lim}$ & 7.21 & 62.4 & 3.91 & 24.2 \\
    \end{tabular}
    \caption{Relative sensitivity of $H_{0}$ measurements for ASKAP/CRACO and MeerKAT coherent observations, using all FRB hosts with spec-zs; adjusting for photo-z errors of $\sigma_z=0.035$; adjusting for a \single\ magnitude limit of $m_r^{\rm lim}$ (of $24.7$ for ASKAPO/CRACO and $27.5$ for MeerTRAP coherent); and including both these latter effects. Sensitivity is measured via the FWHM of likelihoods calculated in Figure~\ref{fig:H0}.}
    \label{tab:H0}
\end{table}

Our results show that our simulated photo-z error of $\sigma_z=0.035$ will have little impact on cosmological parameter estimation. This should not be surprising, given that the $p({\rm DM}|z)$ distribution has a spread of ${\mathcal O}\sim100$\,\pccc, which corresponds to an uncertainty of approximately $\pm 0.1$\,z given the slope of the Macquart (z--DM) relation \citep{Macquart2020}. Hence, an error of $0.035$\,z, added in quadrature, would naively be expected to reduce sensitivity by ${\mathcal O}\sim 6$\% --- very close to the 6.8\% and 3.3\% increases we find for ASKAP/CRACO and MeerTRAP coherent. The effect for MeerTRAP is lower because, due to fluctuations in the cosmological DM, intrinsic variation in the Macquart relation increases with redshift, and this reduces the relative importance of redshift errors. We caution that our results do not include major outliers in photo-zs, which we implicitly assume can be removed by quality cuts. Furthermore, as noted in the Introduction, FRB hosts are not a fully representative sample of field galaxies, so that photo-z estimation for the subcategory of FRB hosts may differ.

A much more significant effect is the reduced fraction of high-redshift FRB host galaxies detectable by Rubin. Our Monte Carlo sampling of the $m_r^{\rm lim}=24.7$ ($27.5$)criteria for ASKAP/CRACO (MeerTRAP coherent) preserved 74\% (78\%) of the original FRBs (expected value 65\% (81\%) from Table~\ref{tab:fractions}), which by simple $N^{0.5}$ scaling, should increase the uncertainty on $H_0$ by 16\% (12\%). However, this cut instead increased the uncertainty by $46.8\%$ ($19.6$\%), implying that the high-redshift hosts have a much greater statistical weight in the analysis. Again, this should not be surprising; in the case of $H_0$, which is determined via the slope of the Macquart relation \citep{Macquart2020,2022MNRAS.516.4862J}, high-redshift FRBs give a larger lever arm.

We also note two potentially biasing effects not simulated here.  
Firstly, it is unclear if there will be a bias in host DM contribution depending on galaxy type. 
While larger and brighter FRB host galaxies are expected to produce larger host DM contributions \citep[e.g.][]{2020ApJ...900..170Z},  a weak anti-correlation is observed within current small samples \citep{2025ApJ...991L..25L}. Whatever the true effect, by preferentially selecting bright galaxies for high-redshift FRBs, optical selection effects may bias the host galaxy DM contribution, and hence skew the $z$--DM relation. Secondly, there is some evidence that strongly repeating FRBs favour dwarf host galaxies, but exhibit a large excess DM, potentially due to ionised media in the vicinity of the progenitor \citep{michilli_extreme_2018,Niu2022}, or ionisation of the star-forming gas reservoir. If this is the case, then optically-limited host observations may be biased against strong repeaters. Such effects suggest that targeted follow-up of FRB host galaxies which cannot be identified by Rubin may still be warranted.

\section{Discussion \& Conclusion}
\label{sec:photozs}

We have investigated the apparent magnitude of FRB host galaxies as a function of redshift using the analysis of \citet{Marnoch2023}, and coupled this to predictions for redshift dependence based on FRB dispersion measure using the \zdm\ code. The resulting model predicts the distribution of FRB host galaxies in $z$--$m_r$ space, and is generally applicable to any optical follow-up observation.

We apply this model to FRBs detected by the coherent observation modes of ASKAP and MeerKAT, and use LSST limiting optical r-band magnitudes corresponding to both a \single\ and \coadds. We find that the expected sensitivity obtained by a \single\ is expected to be sufficient to identify almost all FRB host galaxies out to $z \sim 0.5$, and a majority out to $z \sim 1$, while the \coadds\ will identify essentially all hosts to $z\sim1$ and the majority to $z\sim2$, reaching up to 50\% around cosmic noon ($z\sim2$). Combined with a survey coverage of the entire Southern Hemisphere, LSST will find the majority of host galaxies from FRBs detected by ASKAP and MeerKAT in their coherent observation modes --- and likely the SKA also.

We have also used estimates for the accuracy of Rubin's photometric redshifts, and included this in a simplified simulation to estimate Hubble's constant, $H_0$. We determine that errors introduced by the use of photometric redshifts are expected to be relatively minor --- in the case of ASKAP/CRACO FRB hosts detected in a \single\ Rubin observation, a 6.8\% reduction in statistical precision is expected, while it is only 3.3\% for MeerTRAP FRB hosts detected in co-added images at completion of the LSST. However, missing high-$z$ FRBs will have a greater effect, leading to 47\% and 24\% decreases in precision, respectively.

A potential, even more powerful, application of Rubin data will be field-level inference of the cosmic baryons in the FRB lines of sight, whereby the passage of FRBs through cosmic structures is reconstructed, and the fraction of baryons in circumgalactic (CGM) and intergalactic (IGM) media statistically determined \citep{2022ApJ...928....9L}. The FLIMFLAM survey used extensive spectroscopic redshift measurements in the direction of eight known FRBs to infer cosmological gas fractions (relative to the critical density) of $f_{\rm IGM}=0.59_{-0.10}^{+0.11}$ and $f_{\rm CGM} = 0.55_{-0.29}^{+0.26}$, but was very observationally intensive \citep{FLIMFLAMdr1}, highlighting the need to use non-targeted surveys such as LSST. This approach can also self-consistently break degeneracies with respect to FRBs with anomalously high host DMs \citep{2023ApJ...954L...7L} or uncertain host associations \citep{2020ApJ...903..152H}. However, further work will be required to determine whether Rubin’s photometric redshifts will provide sufficient information for such studies. Combining information from other surveys such as Euclid, Roman, and WAVES may significantly improve this issue. Redshift uncertainties tend to rise at low redshifts, as discussed in \citet{newman}. However, it is much more feasible to obtain spectroscopy of lower redshift galaxies via ground-based spectroscopic follow up or the aforementioned surveys, so we ignore this effect here.

We acknowledge that several aspects of our study are imprecise --- particularly in our treatment of photo-z errors --- and more sophisticated techniques will need to be used when performing future analyses based on the first, and subsequent, annual LSST data releases. Nonetheless, we conclude that the advent of the Vera C.\ Rubin Observatory and its Legacy Survey of Space and Time will be a monumental enabler for identifying the host galaxies of FRBs detected in the Southern Hemisphere, and facilitating subsequent cosmological studies. Furthermore, as the next generation of radio telescopes overlapping with LSST is expected to localise more FRBs than is practical to obtain comprehensive spectroscopic followup, photometric redshifts may become a necessity to truly understand FRB hosts as a population.

\begin{acknowledgement}
Australian access to LSST data is enabled by funding from the Australian Government through an Australian Research Council Linkage Infrastructure, Equipment and Facilities grant LE220100007, as well as the National Collaborative Research Infrastructure Strategy (NCRIS). CRACO was funded through Australian Research Council (ARC) Linkage Infrastructure Equipment, and Facilities grant LE210100107.  RMS acknowledges support through  ARC Discovery Project DP220102305.
\end{acknowledgement}

\paragraph{Funding Statement}

The authors declare no external sources of funding.

\paragraph{Competing Interests}

The authors declare no competing interests.

\paragraph{Data Availability Statement}

The data and results from this work can be obtained from the \zdm\ GitHub codebase \citep{zdm}, at \url{https://github.com/FRBs/zdm}.

\printendnotes

\printbibliography

\end{document}